# Synthesis and physical properties of (Pb$_{0.5}$*M*$_{0.5}$)(Sr,La)$_2$CuO$_z$ (*z*~5; *M* = Fe, Co, Cu, and Zn)

Takumi Nakano [1], Toshihiko Maeda [1,2,*], Takeshi Fujita [1,2], Aichi Yamashita [3,*,†]

[1] *School of Environmental Science and Engineering, Kochi University of Technology, Kami, Kochi 782-8502, Japan.*

[2] *Center for Nanotechnology, Kochi University of Technology, Kami, Kochi 782-8502, Japan.*

[3] *Department of Physics, Tokyo Metropolitan University, 1-1, Minami-osawa, Hachioji 192-0397, Japan.*

**Abstract**

(Pb$_{0.5}$Cu$_{0.5}$)(Sr$_{0.5}$La$_{0.5}$)$_2$CuO$_z$ (abbreviated as (Pb,Cu)-"1-2-0-1") with superconducting transition temperature ($T_c$) of 25 K is a member (*n* = 1) of one of the homologous series of cuprate superconductors, (Pb$^{4+}$,Cu$^{2+}$)(Sr$^{2+}$,*Ln*$^{3+}$)$_2$(Y$^{3+}$,Ca$^{2+}$)$_{n-1}$Cu$^{2+}$$_n$O$^{2-}$$_{2n+3}$ (*n* = 1–4; *Ln*: lanthanoid elements). For the (Pb,Cu)-"1-2-0-1", substitution effects of 3d transition metal elements *M* (*M* = Fe, Co, and Zn) for the Cu site in the (Pb,Cu)–O charge-reservoir layer (labelled as Cu(1)) are systematically investigated. Because Fe, Co and Ni ions exist as divalent or trivalent in ionic crystals, the Sr$^{2+}$/La$^{3+}$ ratio in the (Sr,*Ln*) site is adjusted to satisfy charge neutrality, assuming that they are in a trivalent state. This results in the successful synthesis of new materials with nominal compositions of (Pb$_{0.5}$*M*$_{0.5}$)(Sr$_{0.75}$La$_{0.25}$)$_2$CuO$_z$ (*M* = Fe and Co). This observation suggests that Fe and Co are trivalent in "1-2-0-1". For *M* = Zn, the nominal composition of (Pb$_{0.5}$Zn$_{0.5}$)(Sr$_{0.5}$La$_{0.5}$)$_2$CuO$_z$ was found to yield a nearly single "1-2-0-1" phase. Temperature dependence of electrical resistivity and magnetization were measured, and superconductivity was confirmed only for the case of *M* = Zn with a $T_c$ of 19.7 K. For these three materials, the distribution of Fe, Co and Zn between Cu(1) and another Cu site in the Cu–O$_2$ plane labelled as Cu(2) was investigated employing transmission electron microscopy, which showed that Fe, Co, and Zn occupy both the Cu(1) and Cu(2) sites.

Highlights

・New Pb-based "1-2-0-1" materials of $(Pb_{0.5}M_{0.5})(Sr_{0.75}La_{0.25})_2CuO_z$ ($M$ = Fe and Co) were synthesized by adjusting the Sr/La ratio.

・Microstructural analysis of the "1-2-0-1" cuprate was performed using high-resolution transmission electron microscopy.

・The physical properties of the "1-2-0-1" compounds, such as electrical resistivity and magnetization, were investigated.

[†]Visiting researcher, *Kochi University of Technology, Kami 782-8502, Japan.*

*Corresponding author: Aichi Yamashita, Toshihiko Maeda

E-mail: aichi@tmu.ac.jp

## 1. Introduction

Since the discovery of high-temperature superconductivity in $(La,Ba)_2CuO_4$ [1,2] in 1986, several families of cuprate superconductors have been discovered [3-8]. Their superconducting transition temperature ($T_c$) varies with the number of Cu–$O_2$ superconducting planes ($n$) [9-11] characteristically contained in their crystal structure. For example, $n$ varies from 1 to 4 in $TlBa_2Ca_{n-1}Cu_nO_{2n+3}$, where $T_c$ increases together with $n$ and reaches the highest value of 135 K at $n$ = 3, then decreases [12] to 125 K at $n$ = 4. These crystal structures are schematically drawn in Figure 1. This is likely caused by an imbalance of carriers supplied by the charge reservoir layer (CRL) [11, 12]. For $n$ = 3, the Cu–$O_2$ inner planes (IP) crammed between two Cu–O$_2$ outer planes (OP) are structurally flat and, because of the relatively long distance, are less affected by the CRL situation [11, 13]. Conversely, because there are two IPs for $n$ = 4, charge transfer from CRL to IPs may get relatively lower [13]. This is likely why $T_c$ for $n$ = 4 is lower than that for $n$ = 3. For $n$ = 1 [14-16], because only one Cu–O$_2$ plane is crammed between two CRLs, properties such as composition, valence state, and magnetic state strongly affect the plane's superconductivity. Therefore, research on the "1-2-0-1" ($n$ = 1) is vital for a comprehensive understanding of the role of the CRL on superconductivity, especially on the carrier-doping mechanism.

Presently, two types of the "1-2-0-1" materials of Hg- [14] and Tl-"1-2-0-1" [15] are well known, both of which are difficult to handle because of the higher toxicity of Hg and Tl. In addition, there is Pb–"1-2-0-1", which may have relatively lower toxicity [16-18]. A schematic illustration of the crystal structure of (Pb$_{0.5}$Cu$_{0.5}$)(Sr$_{0.5}$La$_{0.5}$)$_2$CuO$_z$ ((Pb,Cu)-"1-2-0-1") is shown in Figure 2 drawn using VESTA [19]; this material was first reported by Adachi et al. [16]. Its $T_c$ was reported to be approximately 25 K and was found to increase up to 35 K by deoxygenation [20]. Furthermore, (Pb,Cu)-"1-2-0-1" is classified as a self-doped p-type superconductor, and the stoichiometric value of its oxygen content, $z$, equals 5. The mixed valence state of Pb$^{4+}$/Pb$^{2+}$ and oxygen non-stoichiometry are thought to play crucial roles in carrier doping into the Cu–O$_2$ plane.

In this study, the substitution effects of several 3d transition metal elements (Fe, Co, and Zn) on Cu sites in the CRL were investigated for (Pb,Cu)-"1-2-0-1". In addition, some physical properties of the obtained materials, such as electrical resistivity and magnetization, were also investigated, mainly from the standpoint of superconductivity.

## 2. Experimental details

Polycrystalline samples with nominal compositions of (Pb$_{0.5}$$M$$_{0.5}$)(Sr$_{0.5}$La$_{0.5}$)$_2$CuO$_z$ ($M$ = Fe, Co, Ni, Cu, and Zn) and (Pb$_{0.5}$$M$$_{0.5}$)(Sr$_{0.75}$La$_{0.25}$)$_2$CuO$_z$ ($M$ = Fe, Co, and Ni) were prepared using a solid-state reaction method. Commercial PbO (99.9%), CuO (99.9%), SrCO$_3$ (99.99%), La$_2$O$_3$ (99%), Fe$_2$O$_3$ (99.9%), Co$_2$O$_3$ (99.99%), NiO (99.9%), and ZnO (99.999%) powders were mixed and calcined at

800°C for 5 h in air and then cooled to room temperature at a cooling rate of 200°C/h. The calcined powder was remixed, pressed into rectangular bars (approximately 3×3×20 mm$^3$), and sintered at 1000°C for 2 h under flowing oxygen. To obtain a pure "1-2-0-1" phase, the sintering conditions were investigated by using the following sintering conditions: 900, 950, and 1000°C for 2, 4, 8, and 16 h in air or oxygen and cooled down to room temperature at a cooling rate of 60°C/h. The optimized conditions are summarized in the supplemental data with $T_c$ and valence for Pb and $M$ ions (Table S1). Quenching treatment to remove oxygen was done by placing samples on a room-temperature copper plate immediately after annealing at 800°C for 1.5 h in air.

The phase purity was evaluated by powder X-ray diffraction (XRD) using CuKα radiation (Rigaku, SmartLab SE); a D/teX Ultra 250 detector was used for this purpose. Furthermore, the crystal structure was refined using the Rietveld analysis with RIETAN-FP [21].

The microstructures were characterized using transmission electron microscopy (TEM) (JEOL, JEM-ARM200F "NEO ARM", equipped with aberration correctors for the image- and probe-forming lens systems, CEOS GmbH) and energy-dispersive X-ray spectroscopy (EDS) (JEOL, JED-2300T). The TEM and scanning TEM (STEM) experiments were conducted at an accelerating voltage of 200 kV. The temperature-dependence of electrical resistivity and magnetization was measured using a conventional four-probe method with an applied current of 3 mA and a superconducting quantum interference device (SQUID) on an MPMS-3 (Quantum Design, Magnetic Property Measurement System-3) with an applied field of 10 Oe. Both were measured within the temperature range of 2–300 K.

The valence states of Pb and Co were evaluated using X-ray absorption near-edge spectroscopy (XANES; Aichi Synchrotron Radiation Center, BL5S1). As references for $Pb^{2+}$, $Pb^{4+}$, $Co^{2+}$ and $Co^{3+}$, PbO, $PbO_2$, CoO, and $Co_2O_3$ were used, respectively. Data were analyzed using Athena.

## 3. Results and discussion

Figures 3(a) and (b) show the XRD profiles of the as-sintered and quenched samples, respectively.

Detailed results of the Rietveld refinement are summarized in the Supplementary files (Tables S2, S3, and Figures. S1, S2). The space group of (Pb,$M$)-"1-2-0-1" was classified as $P4/mmm$ (No. 123). A single "1-2-0-1" phase for $M$ = Cu and an almost single "1-2-0-1" phase for $M$ = Zn were obtained with the composition $Sr^{2+}/La^{3+}$ = 0.5/0.5, in the (Sr,$Ln$) site, which is the same as the reported values [17, 22, 23]. However, the target materials were not obtained in the case of $M$ = Fe and Co using a ratio of $Sr^{2+}/La^{3+}$ = 0.5/0.5. Taking into account the multi-valence of Fe and Co ions, the basic composition ratio of $Sr^{2+}/La^{3+}$ = 0.5/0.5 was changed to 0.75/0.25, $([Pb^{4+}_{0.5}M^{3+}_{0.5}]^{3.5+})([Sr^{2+}_{0.75}La^{3+}_{0.25}]^{2.25+})_2([Cu^{2+}]^{2+})([O^{2-}_z]^{10.0-})$, which resulted in a successful synthesis of two new kinds of "1-2-0-1" phases with $M$ = Fe and Co. We also attempted to synthesise the "1-2-0-1" with $M$ = Ni; however, only $Sr_5Pb_3(Cu,Ni)O_{12}$ and $(La,Sr)_2(Cu,Ni)O_4$ appeared instead of the "1-2-0-1".

Figures 3(c) and (d) show the lattice constants $a$ and $c$ of the four kinds of (Pb,$M$)-"1-2-0-1". While no significant difference in the $a$-axis length was observed, a significant shrinkage in the $c$-axis length was found for $M$ = Fe and Co compared to that for $M$ = Cu and $M$ = Zn. As described below, the XANES results clearly showed that the Co ion was trivalent, indicating that this shrinkage is due to the smaller ionic radius of $Co^{3+}$ (ionic radius of 0.61 Å for CN = 6) compared to that of $Cu^{2+}$ (0.73 Å) and $Zn^{2+}$ (0.74 Å). Although the information on the valence state of Fe has not yet been obtained, the above successful results strongly support that Fe was also in the trivalent state. Furthermore, the shrinkage of the $a$-axis and elongation of the $c$-axis observed for quenched samples can be attributed to the decrease in oxygen content, as reported previously [24, 25].

Figure 4 shows the results of the EDS with an [100] incident beam using STEM. For $M$ = Cu, each element occupied the crystallographic site described in the nominal composition. However, for $M$ = Fe, Co, and Zn, the nominal compositions were designed assuming that these replaced the Cu(1) site only. The STEM–EDS analysis revealed that the Cu(2) site was also occupied by $M$ ions. Table 1 shows the distribution ratio of $M$ in the Cu(1) and Cu(2) sites determined using the area ratios of the EDS line profiles of Fe, Co, and Zn, which clearly show that Zn preferentially occupied the Cu(1) sites. This seems to be a behavior different from what has been reported for Zn in $YBa_2Cu_3O_z$ [26]. Based on these results, the actual composition was estimated for the samples of $M$ = Fe, Co, and Zn, which are described as $(Pb_{0.5}Cu_{0.3}Fe_{0.2})(Sr_{0.75}La_{0.25})_2(Cu_{0.7}Fe_{0.3})O_z$, $(Pb_{0.5}Cu_{0.25}Co_{0.25})(Sr_{0.75}La_{0.25})_2(Cu_{0.75}Co_{0.25})O_z$, and $(Pb_{0.5}Cu_{0.05}Zn_{0.45})(Sr_{0.5}La_{0.5})_2(Cu_{0.95}Zn_{0.05})O_z$, respectively. We will hereafter refer to the samples as $(Pb_{0.5}M_{0.5})(Sr_{0.5}La_{0.5})_2CuO_z$ for $M$ = Cu and Zn and $(Pb_{0.5}M_{0.5})(Sr_{0.75}La_{0.25})_2CuO_z$ for $M$ = Fe and Co using their nominal compositions, for simplicity.

Table 2 shows the average valence of Pb and Co determined using the XANES results for samples of $M$ = Co and Cu, for which single-phase samples were obtained, together with the values of $T_c$. The average valence of the Co ions was substantially trivalent for both the as-sintered and quenched samples, suggesting that charge compensation by adjusting the $Sr^{2+}/La^{3+}$ ratio was effective. Furthermore, for these samples, the average valence of Pb ions was 3.64+ and 3.54+ for the samples of $M$ = Co and 3.68+ and 3.59+ for the samples of $M$ = Cu, respectively. In both cases, the amount of tetravalent Pb was smaller in the quenched samples. This Pb reduction behavior seems to be in good agreement with a previous report [24].

Figures 5(a)–(f) and 6(a)–(d) show the temperature-dependence of electrical resistivity and magnetization, respectively. For easier understanding, the results were categorized by each $M$ element as follows:

(i) For $M$ = Fe and Co, Figures 5(a) and (b) show the temperature dependence of the resistivity for samples of $M$ = Fe and $M$ = Co, respectively. The results for the as-sintered and quenched samples are presented together, and all four samples exhibit semiconducting behavior. Magnetization measurements down to 2 K also showed that these were nonsuperconducting (Figures 6(a), (b)). A slight divergence between zero field cooling (ZFC) and field cooling (FC) magnetizations is also observed. Both samples exhibited ferromagnetic ordering down to approximately 50 K. Furthermore, a magnetic transition from ferromagnetic to antiferromagnetic ordering was observed for the samples of $M$ = Fe and paramagnetic ordering for those of $M$ = Co. This disappearance of superconductivity could be attributed to the electrical turbulence of the Cu–O$_2$ planes owing to the partial substitution of the magnetic elements for the Cu(2) sites [27].

(ii) For $M$ = Cu, the temperature dependence of electrical resistivity with the metallic and superconducting transition with $T_c^{onset}$ = 27.5 K and 35.5 K was observed for as-sintered and quenched samples, respectively. The as-sintered sample exhibited a slightly higher residual resistivity ratio (*RRR*) of 3.5 compared to that of 2.7 in the quenched sample. The superconducting volume fraction (SVF) estimated from the magnetization was 82.5 and 81.8% for the as-sintered and quenched samples, respectively. The observed $T_c$ values for the as-sintered and quenched samples are consistent with previously reported data [20, 24, 25]. The improvement in $T_c$ by quenching treatment in $M$ = Cu is attributed to the reduction of some Pb ions due to the removal of oxygen, providing additional holes as charge carriers in the Cu–O$_2$ layer [25].

(iii) For $M$ = Zn, the superconducting transition at $T_c$ = 19.7 K was confirmed for the as-sintered sample and $T_c$ = 18.2 K for the quenched sample in both the resistivity and magnetization measurements. Like the case of $M$ = Cu, the as-sintered sample exhibited a slightly higher RRR of 2.9 compared to that of 2.1 in the quenched sample. The SVF values were estimated to be 29.3% and 14.8% for the as-sintered and quenched samples, respectively. However, although only the $T_c$ value has been previously reported [17], this is the first report revealing the whole picture of the temperature dependence of resistivity and magnetization for samples with $M$ = Zn. The decrease in SVF compared to the case of $M$ = Cu could also be due to the electrical turbulence of the Cu–O$_2$ planes induced by Zn substitution. For $M$ = Fe, Co, and Zn, superconductivity is observed only for $M$ = Zn, which may be mainly caused by the relatively small amount of non-magnetic Zn replacing the Cu(2) sites. Unlike the case of $M$ = Cu, the quenching treatment resulted in a slight decrease in $T_c$.

## 4. Conclusions

Herein, we reported the synthesis and physical properties of Pb-based "1-2-0-1" cuprates, (Pb$_{0.5}$$M$$_{0.5}$)(Sr,La)$_2$CuO$_z$, with 3d transition metals $M$ = Fe, Co, Cu, and Zn. Two types of new compounds, $(Pb_{0.5}M_{0.5})(Sr_{0.75}La_{0.25})_2CuO_z$ with $M$ = Fe and Co, were successfully obtained by adjusting the $Sr^{2+}/La^{3+}$ ratio. XRD results showed an almost single "1-2-0-1" phase for $(Pb_{0.5}M_{0.5})(Sr_{0.5}La_{0.5})_2CuO_z$ ($M$ = Cu and Zn) and $(Pb_{0.5}M_{0.5})(Sr_{0.75}La_{0.25})_2CuO_z$ ($M$ = Fe and Co). The valence state of Co in $(Pb_{0.5}M_{0.5})(Sr_{0.75}La_{0.25})_2CuO_z$ ($M$ = Co) was confirmed to be trivalent by XANES measurements. Therefore, adjusting the $Sr^{2+}/La^{3+}$ ratio by the valence of M is considered effective in satisfying charge neutrality. Furthermore, EDS analysis showed that Fe, Co, and Zn ions occupied both the Cu(1) and Cu(2) sites. While the samples with $M$ = Fe and Co did not exhibit a superconducting transition down to 2.0 K, those with $M$ = Cu and Zn were superconductive; $T_c$ was respectively measured to be 27.5 K and 35.5 K in the as-sintered and quenched samples for $M$ = Cu and 19.7 K and 18.2 K for $M$ = Zn, respectively. The discovery of these new (Pb,$M$)-"1-2-0-1" compounds with $M$ = Fe and Co provides an additional pathway for exploring novel (Pb,$M$)-"1-2-0-1" superconductors, as well as investigations of the effects of slight doping of Fe and Co on superconductivity.

**CRediT authorship contribution statement**

Takumi Nakano: Investigation, Writing - Review & Editing. **Toshihiko Maeda**: Conceptualization, Writing - Review & Editing, Supervision. **Aichi Yamashita**: Investigation, Writing - Original Draft, Review & Editing. **Takeshi Fujita**: Investigation, Writing - Review & Editing.

**Declaration of competing interest**

The authors declare that they have no known competing financial interests or personal relationships that could have appeared to influence the work reported in this paper.

**Acknowledgments**

The authors thank Y. Mizuguchi and O. Miura for their support in experiments. This work was partly supported by JSPS KAKENHI (Grant Nos. 21K18834, 22K14480), Thermal and Electric Energy Technology Foundation, Iketani Science and Technology Foundation.

**Supplementary materials**

Supplementary material associated with this article can be found, in the online version, at doi: xxxx

**References**

[1] J.G. Bednorz, and K.A. Müller, Possible highT c superconductivity in the Ba−La−Cu−O system, Z. Phys. B-Cond. Mat. **64** (1986) 189–193, https://link.springer.com/article/10.1007/bf01303701

[2] H. Takagi, S. Uchida, K. Kitazawa, and S. Tanaka, High-$T_c$ Superconductivity of La-Ba-Cu Oxides. II. – Specification of the Superconducting Phase, Jpn. J. Appl. Phys. **26** (1987) L123, DOI: 10.1143/JJAP.26.L123

[3] S. Ideta, K. Takashima, M. Hashimoto, T. Yoshida, A. Fujimori, H. Anzai, T. Fujita, Y. Nakashima, A. Ino, M. Arita, H. Namatame, M. Taniguchi, K. Ono, M. Kubota, D.H. Lu, Z.-X. Shen, K.M. Kojima, and S. Uchida, Enhanced Superconducting Gaps in the Trilayer High-Temperature $Bi_2Sr_2Ca_2Cu_3O_{10+\delta}$ Cuprate Superconductor, Phys. Rev. Lett. **104** (2010) 227001, DOI: 10.1103/PhysRevLett.104.227001

[4] B. Keimer, S. A. Kivelson, M.R. Norman, S. Uchida, and J. Zaanen, From quantum matter to high-temperature superconductivity in copper oxides, Nature **518** (2015) 179–186, https://www.nature.com/articles/nature14165

[5] B. Loret, N. Auvray, Y. Gallais, M. Cazayous, A. Forget, D. Colson, M.-H. Julien, I. Paul, M. Civelli, and A. Sacuto, Intimate link between charge density wave, pseudogap and superconducting energy scales in cuprates, Nat. Phys. **15** (2019) 771–775, DOI: 10.1038/s41567-019-0509-5

[6] B. Loret, A. Forget, J.-B. Moussy, S. Poissonnet, P. Bonnaillie, G. Collin, P. Thuéry, A. Sacuto, and D. Colson, Crystal Growth and Characterization of $HgBa_2Ca_2Cu_3O_{8+\delta}$ Superconductors with the Highest Critical Temperature at Ambient Pressure, Inorg. Chem. **56** (2017) 9396–9399, https://doi.org/10.1021/acs.inorgchem.7b01372

[7] Md. Atikur Rahman, Md. Zahidur Rahaman, and Md. Nurush Samsuddoha, A Review on Cuprate Based Superconducting Materials Including Characteristics and Applications, Am. j. phys. appl. **3** (2015) 39–56, https://www.sciencepublishinggroup.com/journal/paperinfo.aspx?journalid=622&doi=10.11648/j.ajpa.20150302.15

[8] T. Maeda, K. Sakuyama, S. Koriyama, H. Yamauchi, and S. Tanaka, Synthesis and characterization of the superconducting cuprates $(Pb,Cu)Sr_2(Y,Ca)Cu_2O_z$, Phys. Rev. B **43** (1991) 7866, DOI: https://doi.org/10.1103/PhysRevB.43.7866

[9] D. Rybicki, M. Jurkutat, S. Reichardt, C. Kapusta, and J. Haase, Perspective on the phase diagram of cuprate high-temperature superconductors, Nat. Commun. **7** (2016) 11413, DOI: https://doi.org/10.1038/ncomms11413

[10] S. Chakravarty, H.-Y. Kee, and K. Völker, An explanation for a universality of transition temperatures in families of copper oxide superconductors, Nature **428** (2004) 53–55, DOI: https://doi.org/10.1038/nature02348

[11] H. Mukuda, S. Shimizu, A. Iyo, and Y. Kitaoka, High-$T_c$ Superconductivity and Antiferromagnetism in Multilayered Copper Oxides –A New Paradigm of Superconducting Mechanism–, J. Phys. Soc. Jpn. **81** (2012) 011008, https://doi.org/10.1143/JPSJ.81.011008

[12] A. Iyo, Y. Tanaka, Y. Kodama, H. Kito, K. Tokiwa, and T. Watanabe, Synthesis and physical properties of multilayered cuprates, Physica C **445** (2006) 17–22, https://doi.org/10.1016/j.physc.2006.03.067

[13] S. Kunisada, S. Isono, Y. Kohama, S. Sakai, C. Bareille, S. Sakuragi, R. Noguchi, K. Kurokawa, K. Kuroda, Y. Ishida, S. Adachi, R. Sekine, T.K. Kim, C. Cacho, S. Shin, T. Tohyama, K. Tokiwa, and T. Kondo, Observation of small Fermi pockets protected by clean $CuO_2$ sheets of a high-$T_c$ superconductor, Science **369** (2020) 833–838, https://www.science.org/doi/10.1126/science.aay7311

[14] S.N. Putilin, E.V. Antipov, O. Chmaissem, and M. Marezio, Superconductivity at 94 K in $HgBa_2CuO_{4+\delta}$, Nature **362** (1993) 226–228, DOI: https://doi.org/10.1038/362226a0

[15] T. Manako, Y. Shimakawa, Y. Kubo, T. Satoh, and H. Igarashi, Superconductivity of $TlBa_{1+x}La_{1-x}CuO_5$ with 1201 structure, Physica C **158** (1989) 143–147, https://doi.org/10.1016/0921-4534(89)90309-2

[16] S. Adachi, K. Setsune, and K. Wasa, Superconductivity in "1201" Lead Cuprate, Jpn. J. Appl. Phys. **29** (1990) L890, DOI: 10.1143/JJAP.29.L890

[17] T. P. Beals, W.G. Freeman, S.R. Hall, M.R. Harison, and J.M. Parberry, Synthesis and properties of a series of layered copper oxide superconductors with a $(Pb_{0.5}Cd_{0.5})$ rock-salt dopant layer, Physica C **205** (1993) 383–396, https://doi.org/10.1016/0921-4534(93)90406-G

[18] H. Sasakura, Y. Akagi, S. Tsukui, and M. Adachi, New superconducting lead cuprate with 1201 structure, $(Pb_{0.5}B_{0.5})(SrLa)CuO_z$, Mat. Lett. **62** (2008) 4400–4402, https://doi.org/10.1016/j.matlet.2008.07.041

[19] K. Momma, and F. Izumi, VESTA 3 for three-dimensional visualization of crystal, volumetric and morphology data, J. Appl. Crystallogr. **44** (2011) 1272-1276, https://doi.org/10.1107/S0021889811038970

[20] S. Adachi, K. Setsune, and K. Wasa, Annealing Effect of $(Pb/Cu)SrLaCuO_y$ Superconductor, Jpn. J. Appl. Phys. **29** (1990) L2183, DOI: 10.1143/JJAP.29.L2183

[21] F. Izumi, and K. Momma, Three-Dimensional Visualization in Powder Diffraction, Solid State Phenom. **130** (2007) 15–20, DOI: https://doi.org/10.4028/www.scientific.net/SSP.130.15

[22] S. Adachi, K. Setsune, and K. Wasa, Crystal Structure of $(Pb/Cu)SrLaCuO_y$ Superconductor, Jpn. J. Appl. Phys. **29** (1990) L1799, DOI: 10.1143/JJAP.29.L1799

[23] H. Sasakura, K. Nakahigashi, A. Hirose, H. Teraoka, S. Minamigawa, K. Inada, S. Noguchi, and K. Okuda, Preparation of Superconducting 1201 Phase in the $(Pb_{1-y}Cu_y)Sr_{2-x}La_xCuO_z$ System, Jpn. J. Appl. Phys. **29** (1990) L1628, DOI: 10.1143/JJAP.29.L1628

[24] M. Shida, E. Ohshima, M. Kikuchi, M. Nagoshi, and Y. Syono, Valence analysis of Pb and Cu and superconductivity of $(Pb,Cu)(Sr,La)_2CuO_y$, Physica C **268** (1996) 95–99, https://doi.org/10.1016/0921-4534(96)00405-4

[25] N.R. Khasanova, F. Izumi, M. Shida, B.C. Chakoumakos, E. Ohshima, M. Kikuchi, and Y. Syono, Structural disorder and charge transfer in the superconductor $(Pb_{0.5}Cu_{0.5})(Sr_{0.5}La_{0.5})_2CuO_{5+\delta}$, Physica C **269** (1996) 115–123, https://doi.org/10.1016/0921-4534(96)00431-5

[26] E. Takayama-Muromachi, Y. Uchida, and K. Kato, Superconductivity of $YBa_2Cu_{3-x}M_xO_y$ (M = Co, Fe, Ni, Zn), Jpn. J. Appl. Phys. **26** (1987) L2087, DOI: 10.1143/JJAP.26.L2087

[27] G. Xiao, M.Z. Cieplak, J.Q. Xiao, and C.L. Chien, Magnetic pair-breaking effects: Moment formation and critical doping level in superconducting $La_{1.85}Sr_{0.15}Cu_{1-x}A_xO_4$ systems (A = Fe, Co, Ni, Zn, Ga, Al), Phys. Rev. B **42** (1990) 8752, DOI: https://doi.org/10.1103/PhysRevB.42.8752

# Tables

**Table 1.** Distribution ratio and estimated composition of $M$ in $(Pb_{0.5}M_{0.5})(Sr_{0.5}La_{0.5})_2CuO_z$ ($M$ = Zn) and $(Pb_{0.5}M_{0.5})(Sr_{0.75}La_{0.25})_2CuO_z$ ($M$ = Fe, Co).

| Nominal composition | $M$ Element | Distribution ratio of $M$ | | Estimated composition |
|---|---|---|---|---|
| | | (1) site | (2) site | |
| $(Pb_{0.5}Fe_{0.5})(Sr_{0.75}La_{0.25})_2CuO_z$ | Fe | 40 % | 60 % | $(Pb_{0.5}Cu_{0.3}Fe_{0.2})(Sr_{0.75}La_{0.25})_2(Cu_{0.7}Fe_{0.3})O_z$ |
| $(Pb_{0.5}Co_{0.5})(Sr_{0.75}La_{0.25})_2CuO_z$ | Co | 50 % | 50 % | $(Pb_{0.5}Cu_{0.25}Co_{0.25})(Sr_{0.75}La_{0.25})_2(Cu_{0.75}Co_{0.25})O_z$ |
| $(Pb_{0.5}Zn_{0.5})(Sr_{0.5}La_{0.5})_2CuO_z$ | Zn | 90 % | 10 % | $(Pb_{0.5}Cu_{0.05}Zn_{0.45})(Sr_{0.5}La_{0.5})_2(Cu_{0.95}Zn_{0.05})O_z$ |

**Table 2.** $T_c$ and average valence of Pb and $M$ ions. The symbols × and - indicate non-superconducting and unmeasured values, respectively.

| Composition | $T_c$ (K) | | Pb valence | | $M$ valence | |
|---|---|---|---|---|---|---|
| | As-sintered | Quenched | As-sintered | Quenched | As-sintered | Quenched |
| $(Pb_{0.5}Cu_{0.3}Fe_{0.2})(Sr_{0.75}La_{0.25})_2(Cu_{0.7}Fe_{0.3})O_z$ | × | × | - | - | - | - |
| $(Pb_{0.5}Cu_{0.25}Co_{0.25})(Sr_{0.75}La_{0.25})_2(Cu_{0.75}Co_{0.25})O_z$ | × | × | 3.64 | 3.54 | 3.0 | 2.95 |
| $(Pb_{0.5}Cu_{0.5})(Sr_{0.5}La_{0.5})_2CuO_z$ | 27.5 | 35.5 | 3.68 | 3.59 | / | / |
| $(Pb_{0.5}Cu_{0.05}Zn_{0.45})(Sr_{0.5}La_{0.5})_2(Cu_{0.95}Zn_{0.05})O_z$ | 19.7 | 18.2 | - | - | - | - |

**Figures and figure captions**

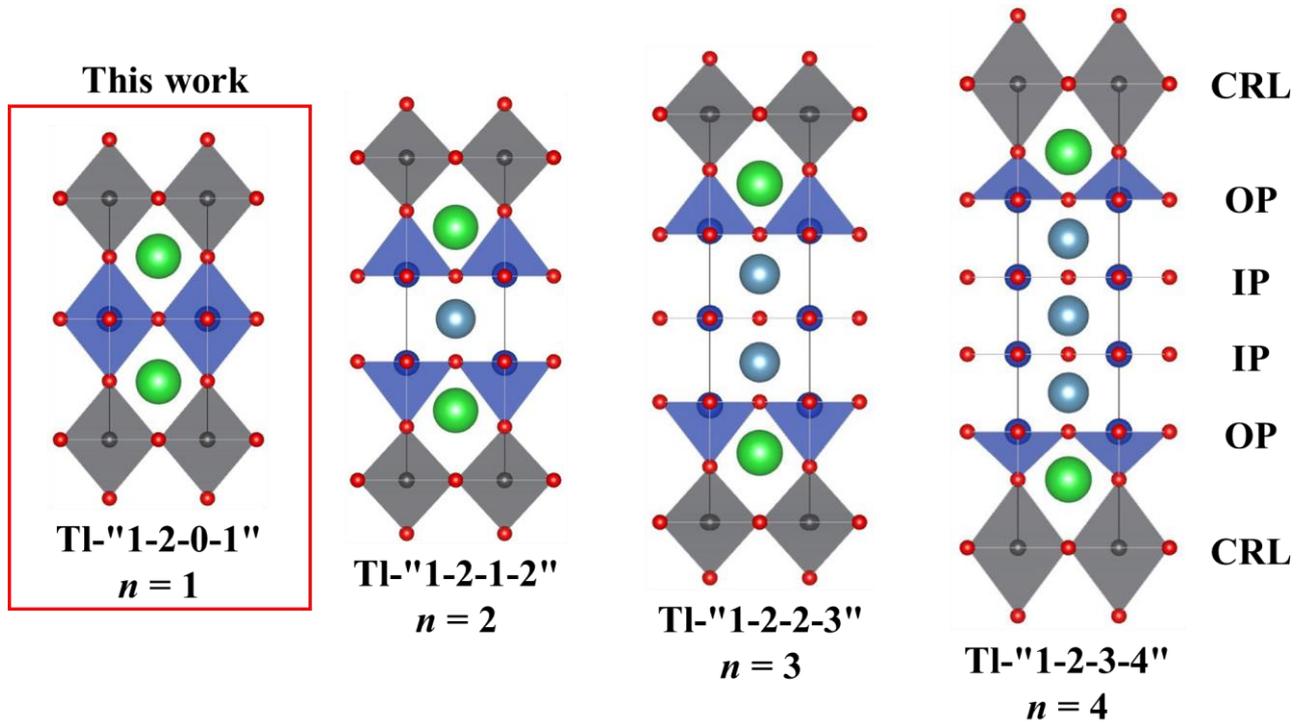

**Fig. 1.** Schematic images of TlBa$_2$Ca$_{n-1}$Cu$_n$O$_{2n+3}$ with $n$ = 1, 2, 3, 4.

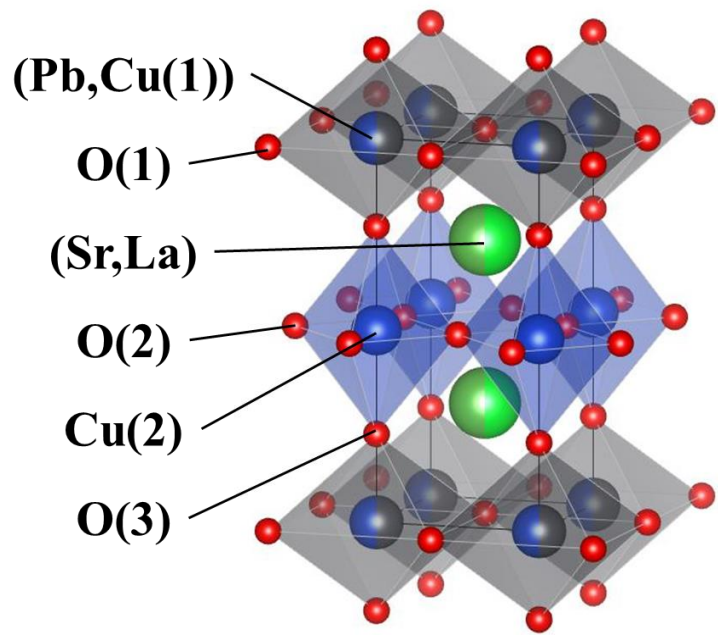

**Fig. 2.** Schematic image of crystal structure of (Pb$_{0.5}$Cu$_{0.5}$)(Sr$_{0.5}$La$_{0.5}$)$_2$CuO$_z$.

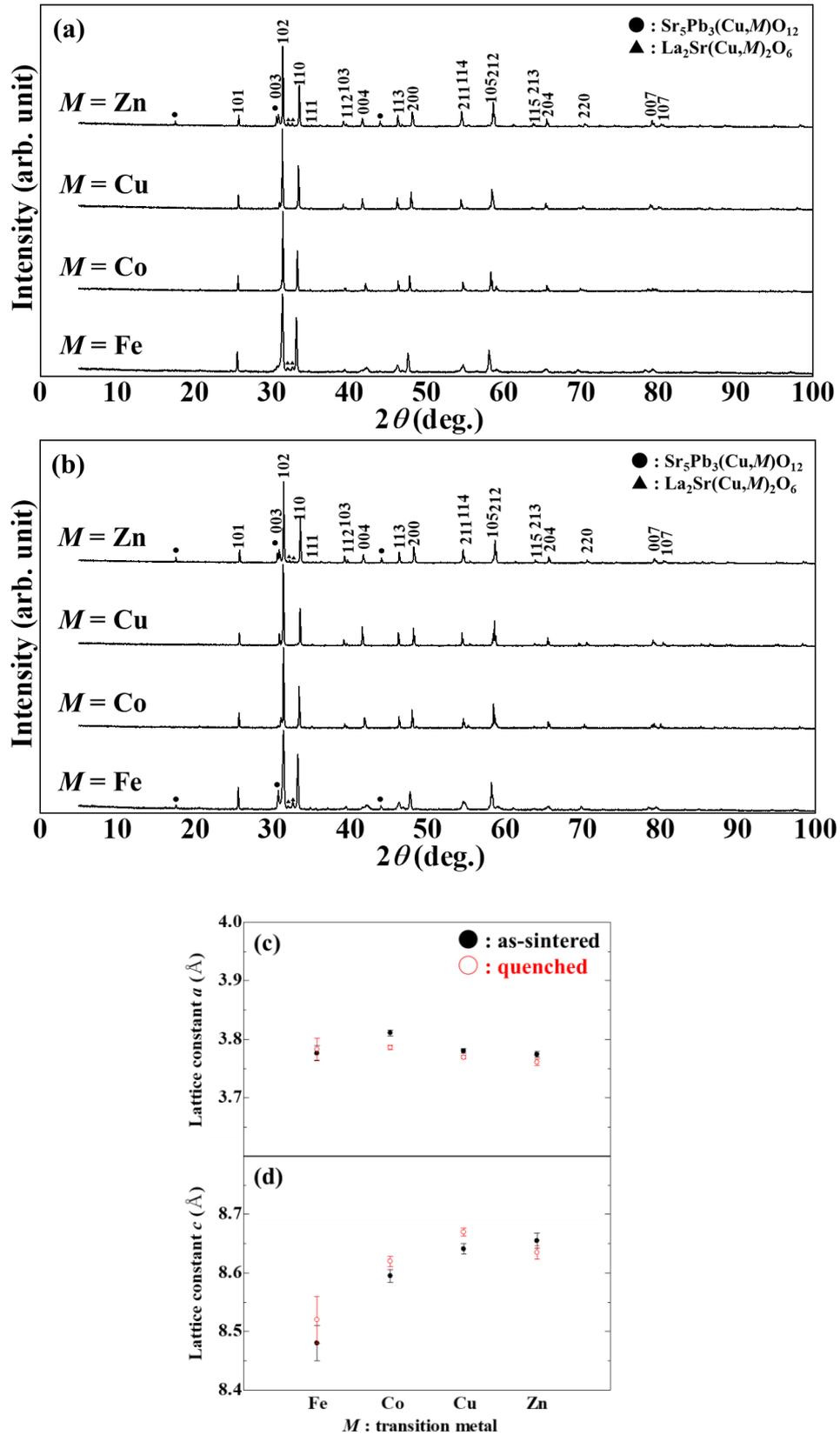

**Fig. 3.** Powder X-ray diffraction patterns of **(a)** as-sintered and **(b)** quenched samples in $(Pb_{0.5}M_{0.5})(Sr_{0.5}La_{0.5})_2CuO_z$ ($M$ = Cu, Zn) and $(Pb_{0.5}M_{0.5})(Sr_{0.75}La_{0.25})_2CuO_z$ ($M$ = Fe, Co). **(c)**, **(d)** show the relationship between lattice constants of the $a$- and $c$-axes and 3d transition metals.

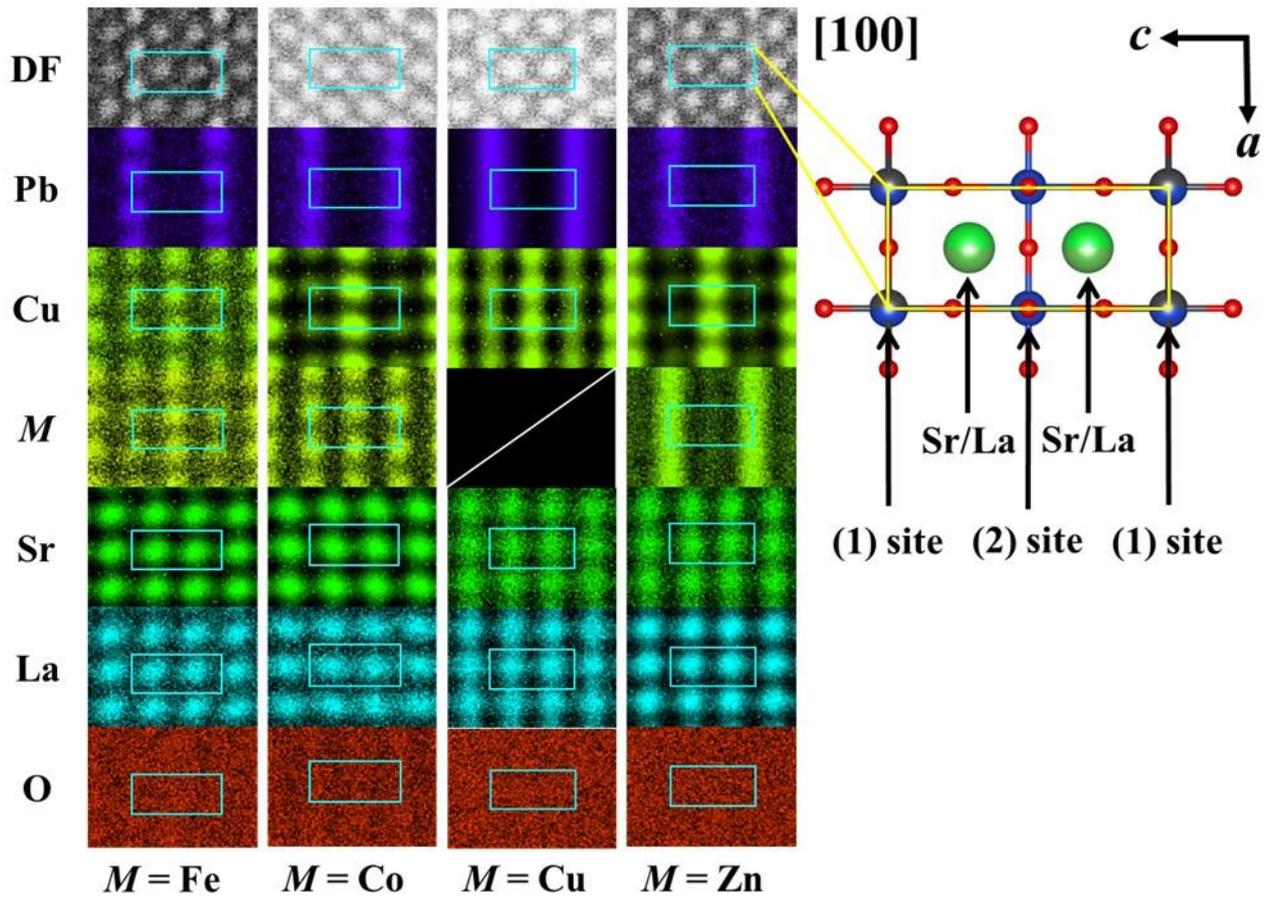

**Fig. 4.** STEM-EDS mapping images for $(Pb_{0.5}M_{0.5})(Sr_{0.5}La_{0.5})_2CuO_z$ ($M$ = Cu, Zn) and $(Pb_{0.5}M_{0.5})(Sr_{0.75}La_{0.25})_2CuO_z$ ($M$ = Fe, Co).

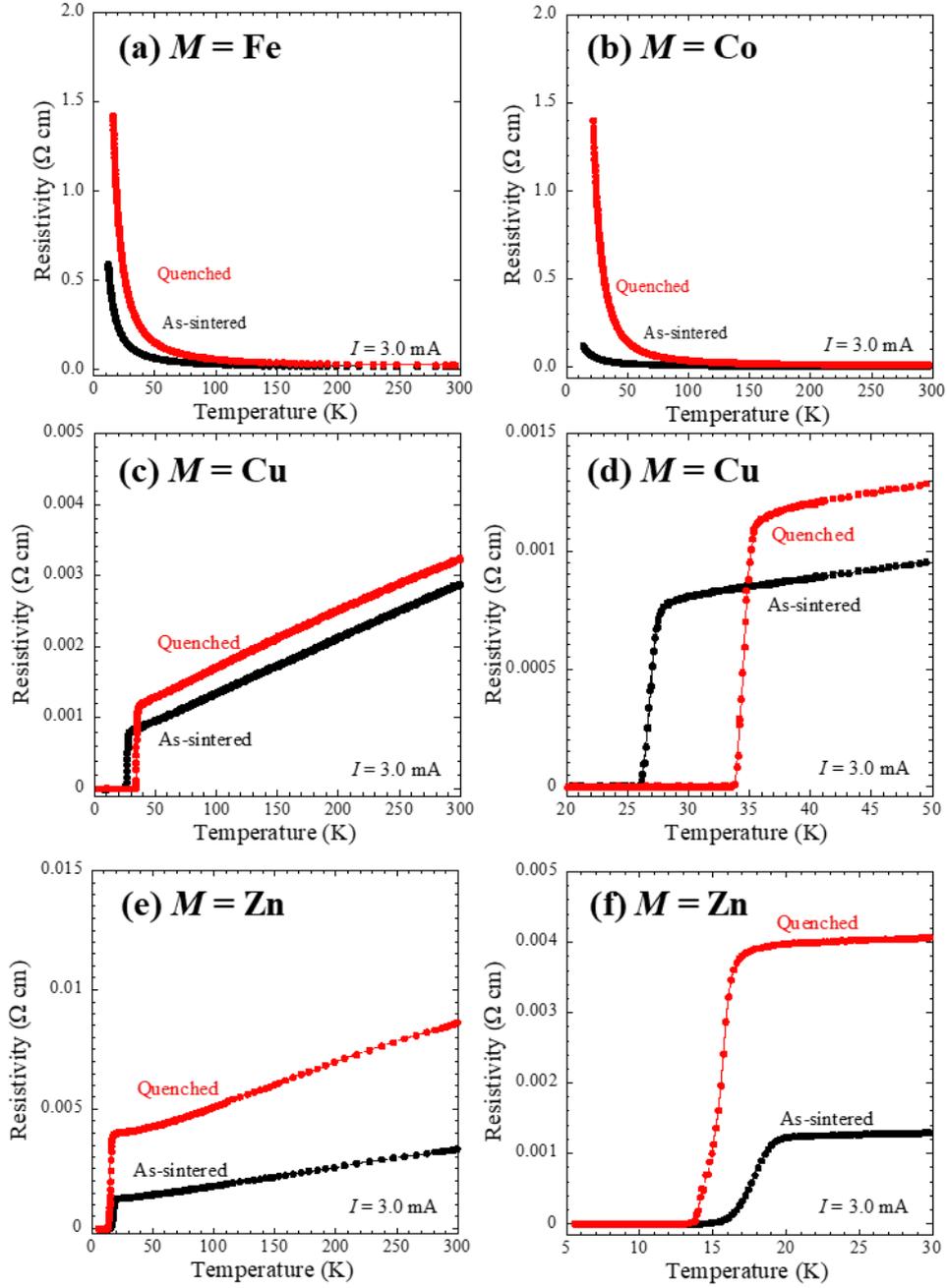

**Fig. 5.** Temperature dependences of electrical resistivity for $(Pb_{0.5}M_{0.5})(Sr_{0.5}La_{0.5})_2CuO_z$ ($M$ = Cu, Zn) and $(Pb_{0.5}M_{0.5})(Sr_{0.75}La_{0.25})_2CuO_z$ ($M$ = Fe, Co).
**(a)** $M$ = Fe, **(b)** $M$ = Co, **(c)** and **(d)** $M$ = Cu, **(e)** and **(f)** $M$ = Zn.

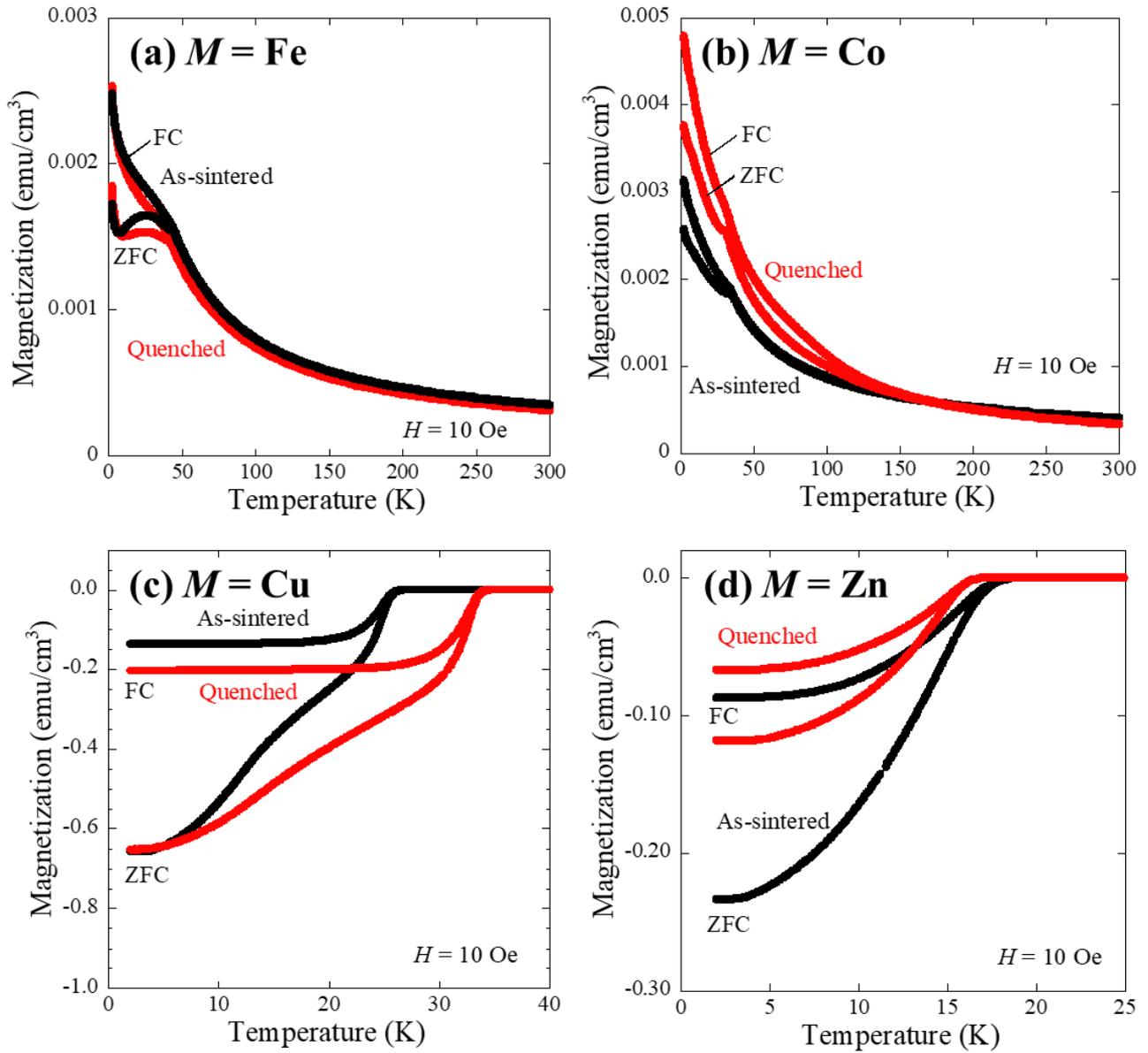

**Fig. 6.** Temperature dependences of magnetization for $(Pb_{0.5}M_{0.5})(Sr_{0.5}La_{0.5})_2CuO_z$ ($M$ = Cu, Zn) and $(Pb_{0.5}M_{0.5})(Sr_{0.75}La_{0.25})_2CuO_z$ ($M$ = Fe, Co).
**(a)** $M$ = Fe, **(b)** $M$ = Co, **(c)** $M$ = Cu and **(d)** $M$ = Zn.

# Supplemental data

Synthesis and physical properties of

$(Pb_{0.5}M_{0.5})(Sr,La)_2CuO_z$ ($z\sim5$; $M$ = Fe, Co, Cu, and Zn)

Takumi Nakano [1], Toshihiko Maeda [1,2,*], Takeshi Fujita [1,2], Aichi Yamashita [3,*,†]

[1] *School of Environmental Science and Engineering*, *Kochi University of Technology*, *Kami, Kochi 782-8502, Japan.*

[2] *Center for Nanotechnology*, *Kochi University of Technology*, *Kami, Kochi 782-8502, Japan.*

[3] *Department of Physics*, *Tokyo Metropolitan University*, *1-1, Minami-osawa, Hachioji 192-0397, Japan.*

[†]Visiting researcher, *Kochi University of Technology, Kami 782-8502, Japan.*

*Corresponding author: Aichi Yamashita, Toshihiko Maeda

E-mail: aichi@tmu.ac.jp

# Tables and Figures

**Table S1.** Table 1, 2 and sintering conditions are shown together.

| Nominal composition | $M$ Element | Distribution rate of $M$ | | Estimated composition | $T_c$ (K) | | Pb valence | | $M$ valence | | Sintering condition | | | |
|---|---|---|---|---|---|---|---|---|---|---|---|---|---|---|
| | | (1) site | (2) site | | As-sintered | Quenched | As-sintered | Quenched | As-sintered | Quenched | Tempelature (°C) | Period (h) | Cooling Rate (°C/h) | Gas |
| $(Pb_{0.5}Fe_{0.5})(Sr_{0.75}La_{0.25})_2CuO_z$ | Fe | 40 % | 60 % | $(Pb_{0.5}Cu_{0.3}Fe_{0.2})(Sr_{0.75}La_{0.25})_2(Cu_{0.7}Fe_{0.3})O_z$ | × | × | - | - | - | - | 950 | 2 | 60 | Air |
| $(Pb_{0.5}Co_{0.5})(Sr_{0.75}La_{0.25})_2CuO_z$ | Co | 50 % | 50 % | $(Pb_{0.5}Cu_{0.25}Co_{0.25})(Sr_{0.75}La_{0.25})_2(Cu_{0.75}Co_{0.25})O_z$ | × | × | 3.64 | 3.54 | 3.0 | 2.95 | 1000 | 2 | 60 | $O_2$ |
| $(Pb_{0.5}Cu_{0.5})(Sr_{0.5}La_{0.5})_2CuO_z$ | Cu | | | | 27.5 | 35.5 | 3.68 | 3.59 | | | 950 | 2 | 60 | $O_2$ |
| $(Pb_{0.5}Zn_{0.5})(Sr_{0.5}La_{0.5})_2CuO_z$ | Zn | 90 % | 10 % | $(Pb_{0.5}Cu_{0.05}Zn_{0.45})(Sr_{0.5}La_{0.5})_2(Cu_{0.95}Zn_{0.05})O_z$ | 19.7 | 18.2 | - | - | - | - | 1000 | 2 | 60 | $O_2$ |

**Table S2.** Structural parameters of as-sintered samples $(Pb_{0.5}M_{0.5})(Sr_{0.5}La_{0.5})_2CuO_z$ ($M$ = Cu, Zn) and $(Pb_{0.5}M_{0.5})(Sr_{0.75}La_{0.25})_2CuO_z$ ($M$ = Fe, Co) refined by Rietveld analysis.
**(a)** $M$ = Fe, **(b)** $M$ = Co, **(c)** $M$ = Cu and **(d)** $M$ = Zn.

(a)

| Atom | Site | $x$ | $y$ | $z$ | $B$ (Å) | $g$ |
|---|---|---|---|---|---|---|
| Pb | 1a | 0 | 0 | 0 | 1/2 | 1/2 |
| Cu1 | 1a | 0 | 0 | 0 | 1/2 | 3/10 |
| Fe1 | 1a | 0 | 0 | 0 | 1/2 | 1/5 |
| Sr | 2h | 1/2 | 1/2 | 0.2875(4) | 1/2 | 3/4 |
| La | 2h | 1/2 | 1/2 | 0.2875(4) | 1/2 | 1/4 |
| Cu2 | 1b | 0 | 0 | 1/2 | 1/2 | 7/10 |
| Fe2 | 1b | 0 | 0 | 1/2 | 1/2 | 3/10 |
| O1 | 1c | 1/2 | 1/2 | 0 | 1 | 1 |
| O2 | 2e | 0 | 1/2 | 1/2 | 1 | 1 |
| O3 | 2g | 0 | 0 | 0.113(3) | 1 | 1 |

(b)

| Atom | Site | $x$ | $y$ | $z$ | $B$ (Å) | $g$ |
|---|---|---|---|---|---|---|
| Pb | 1a | 0 | 0 | 0 | 1/2 | 1/2 |
| Cu1 | 1a | 0 | 0 | 0 | 1/2 | 1/4 |
| Co1 | 1a | 0 | 0 | 0 | 1/2 | 1/4 |
| Sr | 2h | 1/2 | 1/2 | 0.2904(3) | 1/2 | 3/4 |
| La | 2h | 1/2 | 1/2 | 0.2904(3) | 1/2 | 1/4 |
| Cu2 | 1b | 0 | 0 | 1/2 | 1/2 | 3/4 |
| Co2 | 1b | 0 | 0 | 1/2 | 1/2 | 1/4 |
| O1 | 1c | 1/2 | 1/2 | 0 | 1 | 1 |
| O2 | 2e | 0 | 1/2 | 1/2 | 1 | 1 |
| O3 | 2g | 0 | 0 | 0.207(2) | 1 | 1 |

(c)

| Atom | Site | $x$ | $y$ | $z$ | $B$ (Å) | $g$ |
|---|---|---|---|---|---|---|
| Pb | 1a | 0 | 0 | 0 | 1/2 | 1/2 |
| Cu1 | 1a | 0 | 0 | 0 | 1/2 | 1/2 |
| Sr | 2h | 1/2 | 1/2 | 0.2889(2) | 1/2 | 1/2 |
| La | 2h | 1/2 | 1/2 | 0.2889(2) | 1/2 | 1/2 |
| Cu2 | 1b | 0 | 0 | 1/2 | 1/2 | 1 |
| O1 | 1c | 1/2 | 1/2 | 0 | 1 | 1 |
| O2 | 2e | 0 | 1/2 | 1/2 | 1 | 1 |
| O3 | 2g | 0 | 0 | 0.2091(18) | 1 | 1 |

(d)

| Atom | Site | $x$ | $y$ | $z$ | $B$ (Å) | $g$ |
|---|---|---|---|---|---|---|
| Pb | 1a | 0 | 0 | 0 | 1/2 | 1/2 |
| Cu1 | 1a | 0 | 0 | 0 | 1/2 | 1/20 |
| Zn1 | 1a | 0 | 0 | 0 | 1/2 | 9/20 |
| Sr | 2h | 1/2 | 1/2 | 0.2880(3) | 1/2 | 1/2 |
| La | 2h | 1/2 | 1/2 | 0.2880(3) | 1/2 | 1/2 |
| Cu2 | 1b | 0 | 0 | 1/2 | 1/2 | 19/20 |
| Zn2 | 1b | 0 | 0 | 1/2 | 1/2 | 1/20 |
| O1 | 1c | 1/2 | 1/2 | 0 | 1 | 1 |
| O2 | 2e | 0 | 1/2 | 1/2 | 1 | 1 |
| O3 | 2g | 0 | 0 | 0.225(2) | 1 | 1 |

**Table S3.** Structural parameters of quenched samples $(Pb_{0.5}M_{0.5})(Sr_{0.5}La_{0.5})_2CuO_z$ ($M$ = Cu, Zn) and $(Pb_{0.5}M_{0.5})(Sr_{0.75}La_{0.25})_2CuO_z$ ($M$ = Fe, Co) refined by Rietveld analysis.
**(a)** $M$ = Fe, **(b)** $M$ = Co, **(c)** $M$ = Cu and **(d)** $M$ = Zn.

(a)

| Atom | Site | x | y | z | B (Å) | g |
|---|---|---|---|---|---|---|
| Pb | 1a | 0 | 0 | 0 | 1/2 | 1/2 |
| Cu1 | 1a | 0 | 0 | 0 | 1/2 | 3/10 |
| Fe1 | 1a | 0 | 0 | 0 | 1/2 | 1/5 |
| Sr | 2h | 1/2 | 1/2 | 0.2889(4) | 1/2 | 3/4 |
| La | 2h | 1/2 | 1/2 | 0.2889(4) | 1/2 | 1/4 |
| Cu2 | 1b | 0 | 0 | 1/2 | 1/2 | 7/10 |
| Fe2 | 1b | 0 | 0 | 1/2 | 1/2 | 3/10 |
| O1 | 1c | 1/2 | 1/2 | 0 | 1 | 1 |
| O2 | 2e | 0 | 1/2 | 1/2 | 1 | 1 |
| O3 | 2g | 0 | 0 | 0.233(3) | 1 | 1 |

(b)

| Atom | Site | x | y | z | B (Å) | g |
|---|---|---|---|---|---|---|
| Pb | 1a | 0 | 0 | 0 | 1/2 | 1/2 |
| Cu1 | 1a | 0 | 0 | 0 | 1/2 | 1/4 |
| Co1 | 1a | 0 | 0 | 0 | 1/2 | 1/4 |
| Sr | 2h | 1/2 | 1/2 | 0.2910(3) | 1/2 | 3/4 |
| La | 2h | 1/2 | 1/2 | 0.2910(3) | 1/2 | 1/4 |
| Cu2 | 1b | 0 | 0 | 1/2 | 1/2 | 3/4 |
| Co2 | 1b | 0 | 0 | 1/2 | 1/2 | 1/4 |
| O1 | 1c | 1/2 | 1/2 | 0 | 1 | 1 |
| O2 | 2e | 0 | 1/2 | 1/2 | 1 | 1 |
| O3 | 2g | 0 | 0 | 0.2282(18) | 1 | 1 |

(c)

| Atom | Site | x | y | z | B (Å) | g |
|---|---|---|---|---|---|---|
| Pb | 1a | 0 | 0 | 0 | 1/2 | 1/2 |
| Cu1 | 1a | 0 | 0 | 0 | 1/2 | 1/2 |
| Sr | 2h | 1/2 | 1/2 | 0.2892(2) | 1/2 | 1/2 |
| La | 2h | 1/2 | 1/2 | 0.2892(2) | 1/2 | 1/2 |
| Cu2 | 1b | 0 | 0 | 1/2 | 1/2 | 1 |
| O1 | 1c | 1/2 | 1/2 | 0 | 1 | 1 |
| O2 | 2e | 0 | 1/2 | 1/2 | 1 | 1 |
| O3 | 2g | 0 | 0 | 0.2279(16) | 1 | 1 |

(d)

| Atom | Site | x | y | z | B (Å) | g |
|---|---|---|---|---|---|---|
| Pb | 1a | 0 | 0 | 0 | 1/2 | 1/2 |
| Cu1 | 1a | 0 | 0 | 0 | 1/2 | 1/20 |
| Zn1 | 1a | 0 | 0 | 0 | 1/2 | 9/20 |
| Sr | 2h | 1/2 | 1/2 | 0.2891(3) | 1/2 | 1/2 |
| La | 2h | 1/2 | 1/2 | 0.2891(3) | 1/2 | 1/2 |
| Cu2 | 1b | 0 | 0 | 1/2 | 1/2 | 19/20 |
| Zn2 | 1b | 0 | 0 | 1/2 | 1/2 | 1/20 |
| O1 | 1c | 1/2 | 1/2 | 0 | 1 | 1 |
| O2 | 2e | 0 | 1/2 | 1/2 | 1 | 1 |
| O3 | 2g | 0 | 0 | 0.226(2) | 1 | 1 |

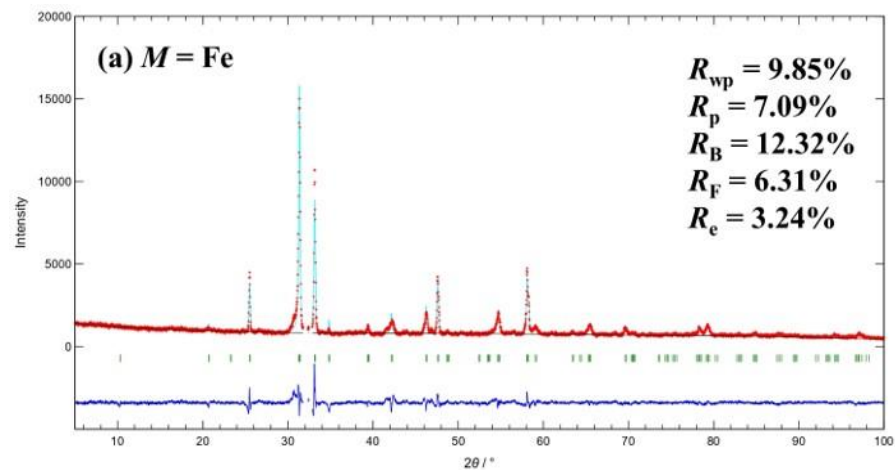
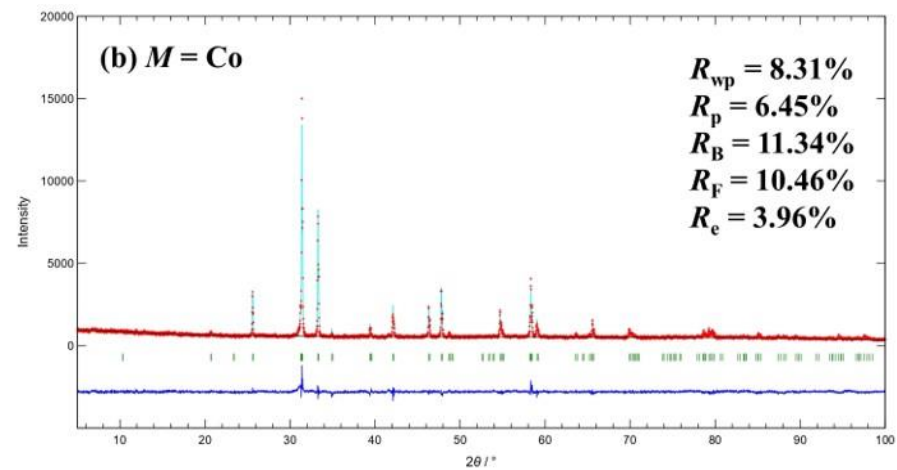
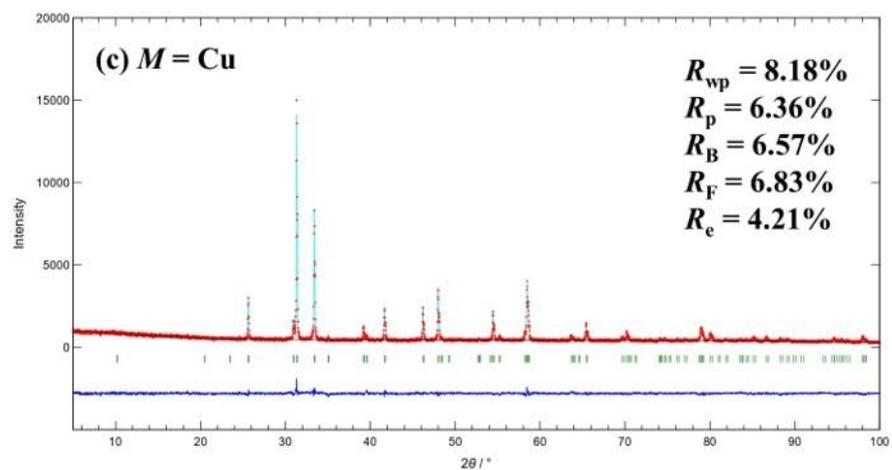
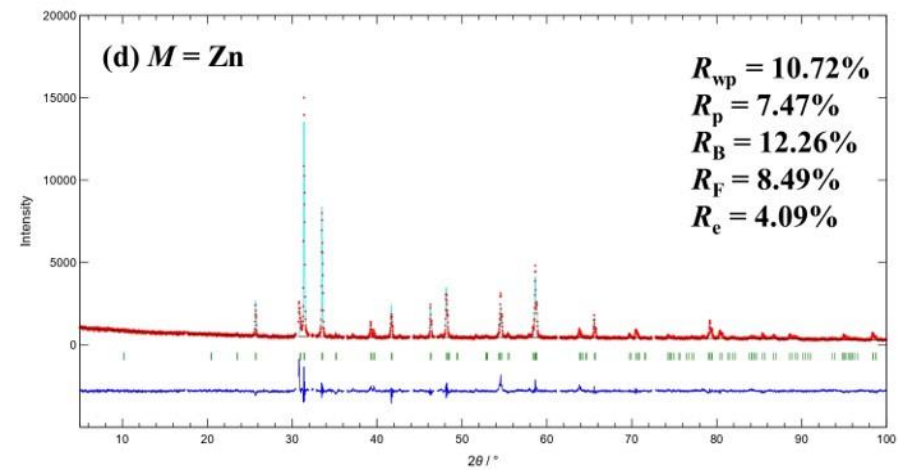

**Fig. S1.** Rietveld analysis results of as-sintered samples $(Pb_{0.5}M_{0.5})(Sr_{0.5}La_{0.5})_2CuO_z$ ($M$ = Cu, Zn) and $(Pb_{0.5}M_{0.5})(Sr_{0.75}La_{0.25})_2CuO_z$ ($M$ = Fe, Co).
**(a)** $M$ = Fe, **(b)** $M$ = Co, **(c)** $M$ = Cu and **(d)** $M$ = Zn.

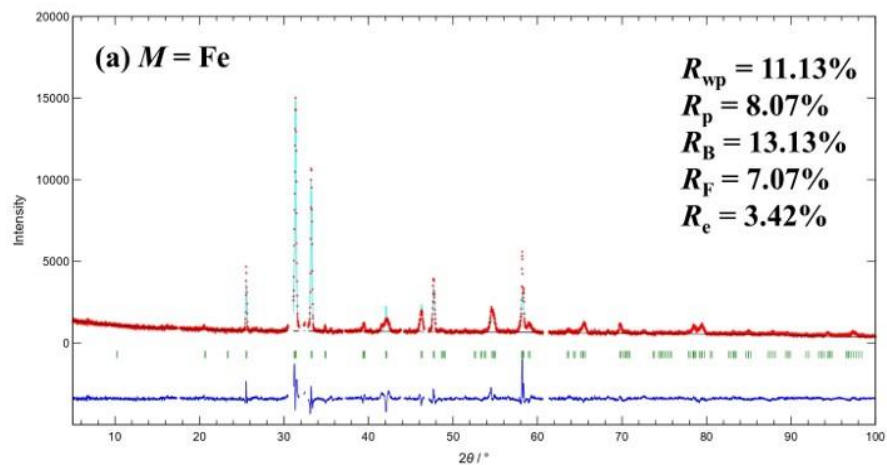
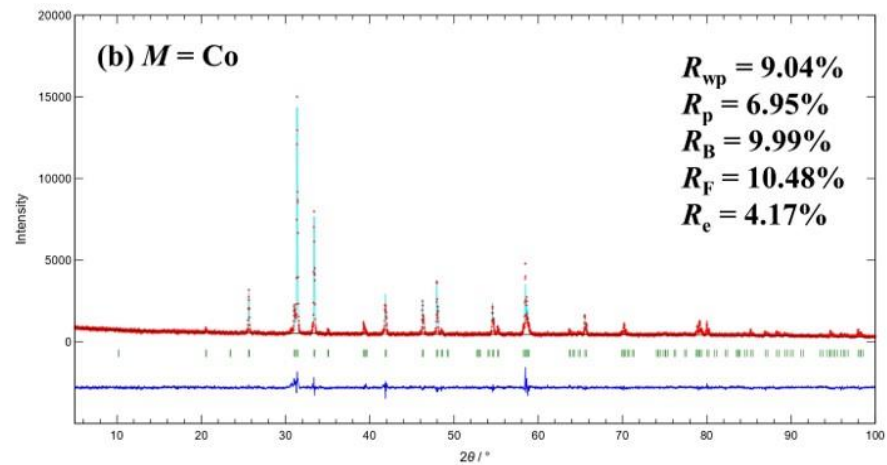
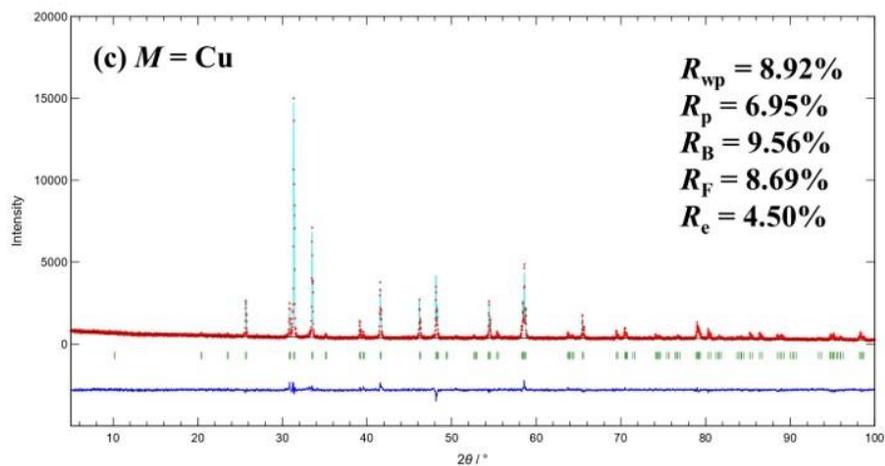
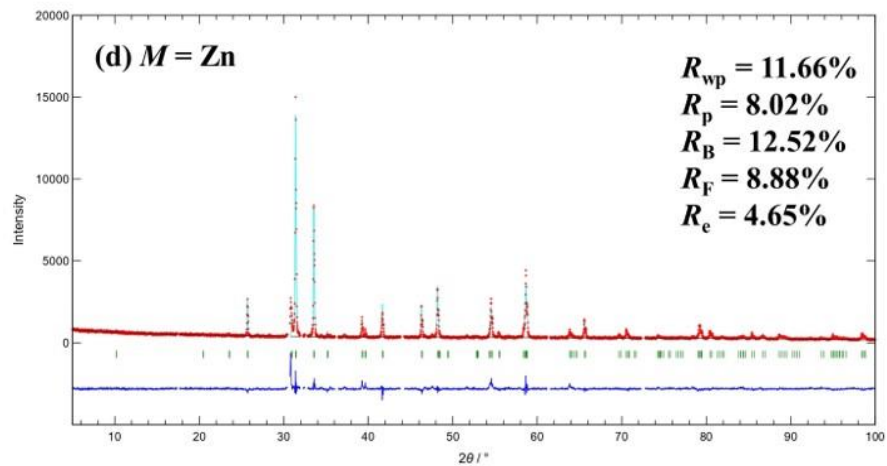

**Fig. S2.** Rietveld analysis results of quenched samples $(Pb_{0.5}M_{0.5})(Sr_{0.5}La_{0.5})_2CuO_z$ ($M$ = Cu, Zn) and $(Pb_{0.5}M_{0.5})(Sr_{0.75}La_{0.25})_2CuO_z$ ($M$ = Fe, Co). **(a)** $M$ = Fe, **(b)** $M$ = Co, **(c)** $M$ = Cu and **(d)** $M$ = Zn.

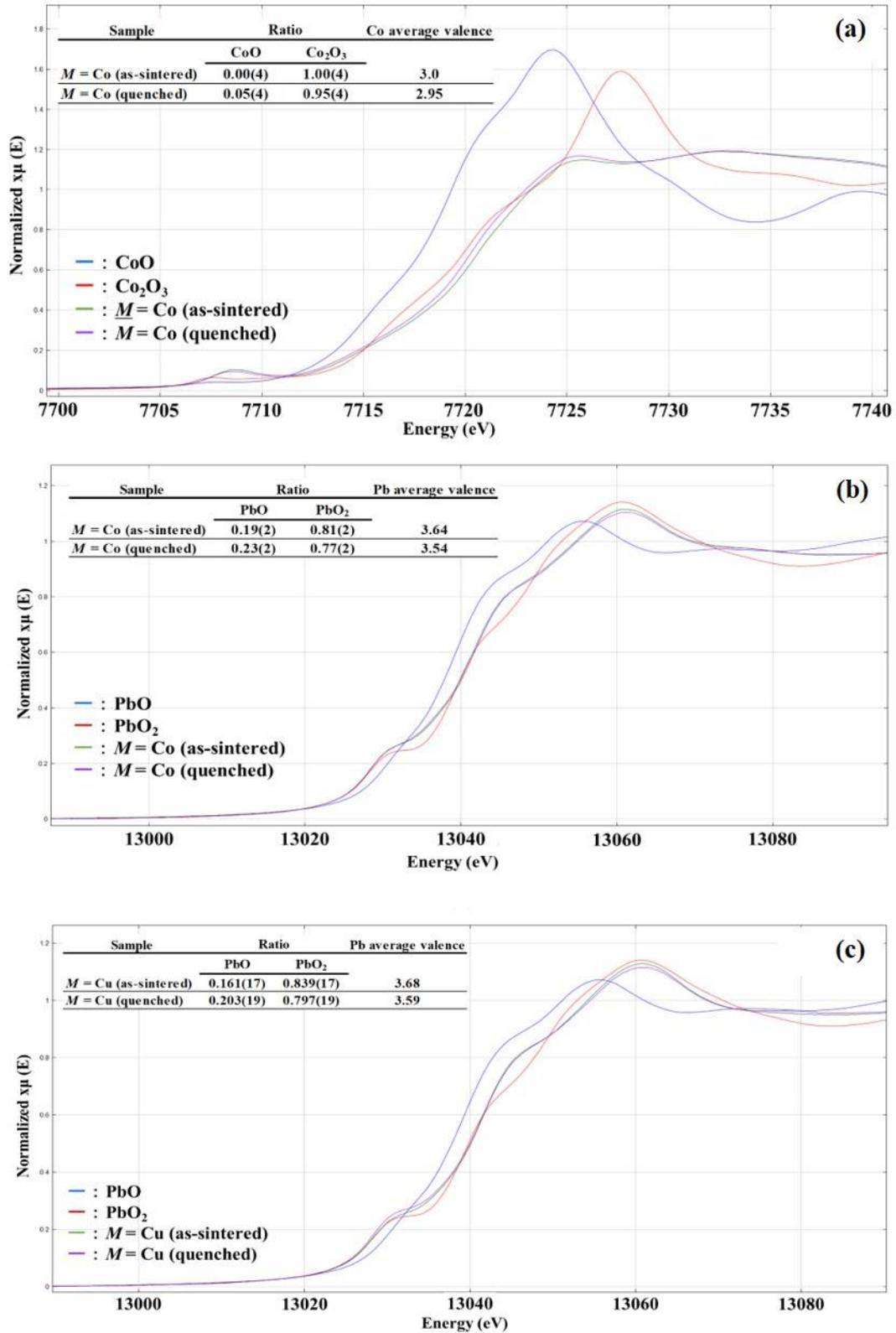

**Fig. S3.** XANES spectra of $(Pb_{0.5}M_{0.5})(Sr_{0.5}La_{0.5})_2CuO_z$ ($M$ = Cu) and $(Pb_{0.5}M_{0.5})(Sr_{0.75}La_{0.25})_2CuO_z$ ($M$ = Co) and average valence of Co and Pb.
**(a)** Spectra near the K absorption edge of Co in $M$ = Co. **(b)**, **(c)** Spectra near the L3 absorption edge of Pb in $M$ = Co and Cu, respectively.

(a)

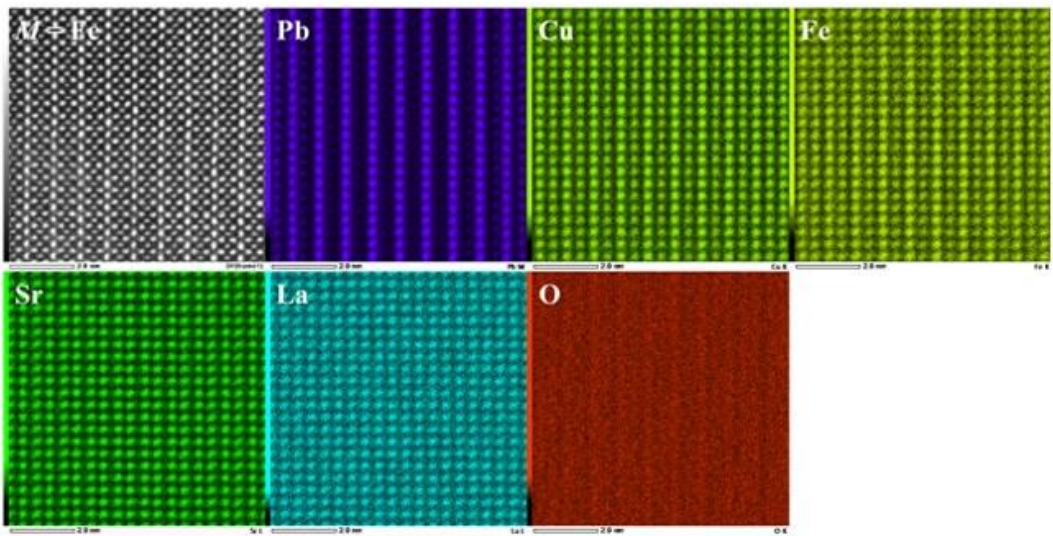

(b)

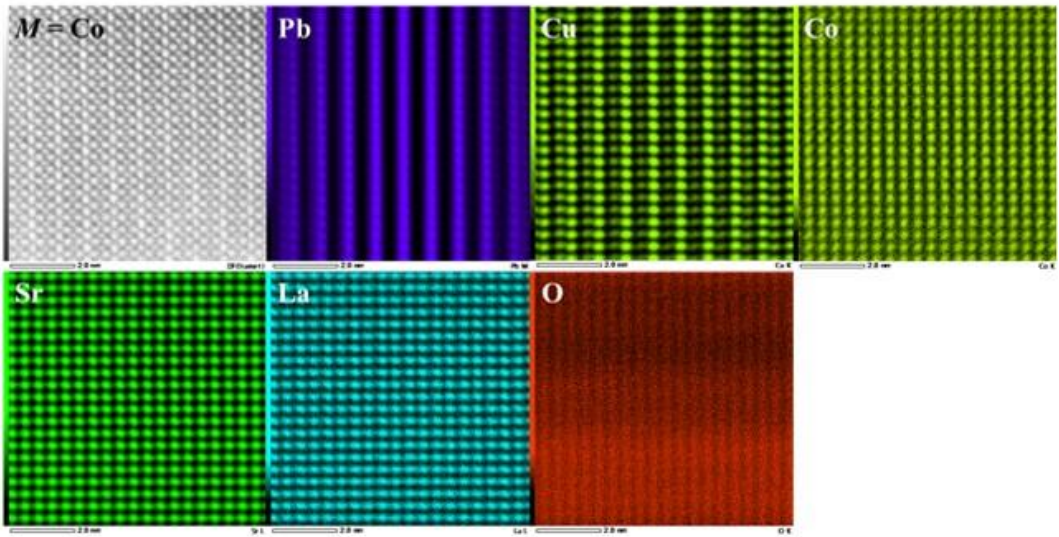

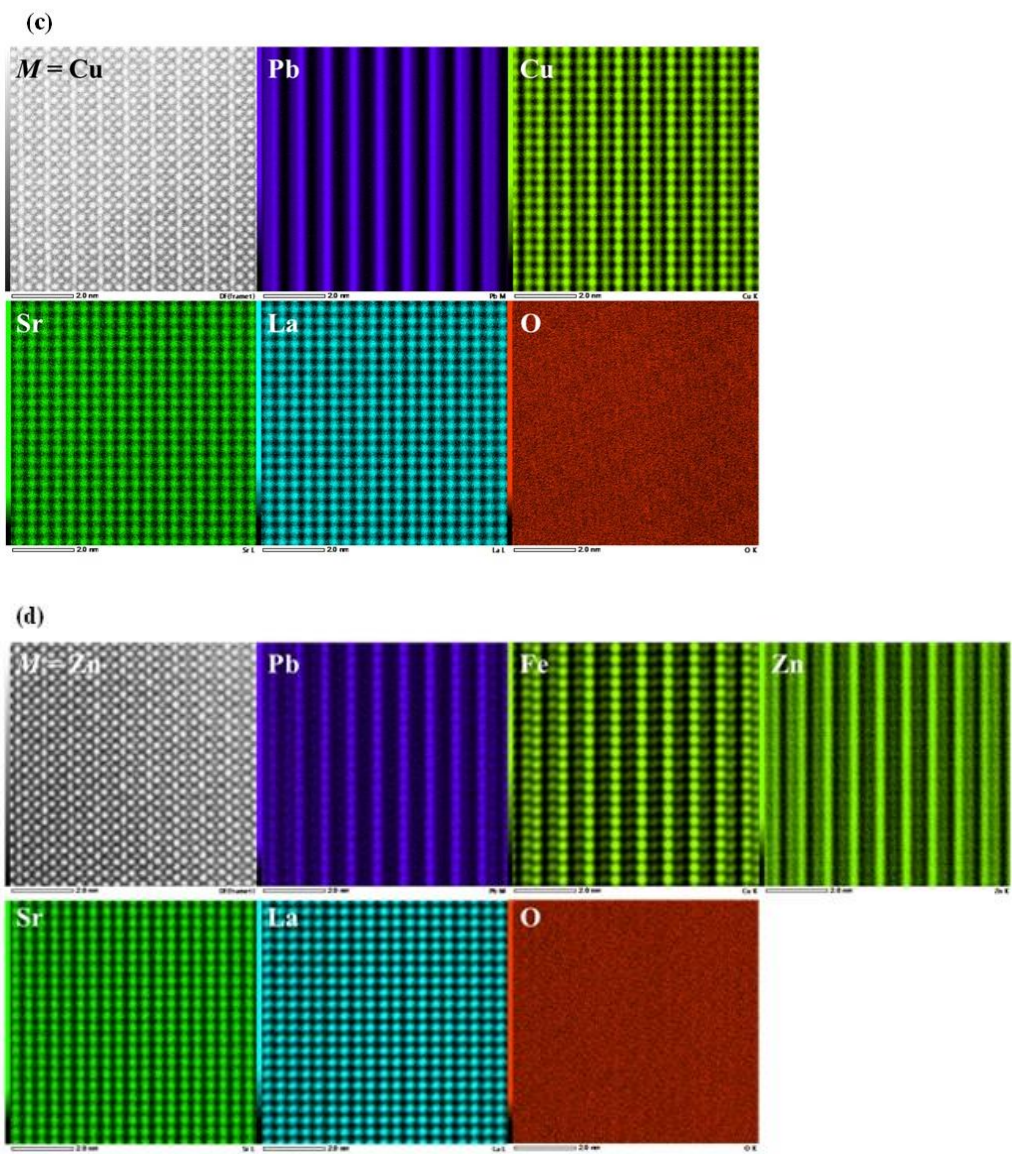

**Fig. S4.** STEM-EDS mapping images for $(Pb_{0.5}M_{0.5})(Sr_{0.5}La_{0.5})_2CuO_z$ ($M$ = Cu, Zn) and $(Pb_{0.5}M_{0.5})(Sr_{0.75}La_{0.25})_2CuO_z$ ($M$ = Fe, Co).
**(a)** $M$ = Fe, **(b)** $M$ = Co, **(c)** $M$ = Cu and **(d)** $M$ = Zn.